\documentclass[a4page]{article}

\usepackage[a4paper, total={6in, 8in}]{geometry}
\usepackage[noblocks]{authblk}
\usepackage{hyperref}
\usepackage{graphicx}

\makeatletter
\renewcommand\AB@affilsepx{, \protect\Affilfont}
\makeatother

\begin{document}

\title{Impressions of the GDMC AI Settlement Generation Challenge in Minecraft\\[1em]
\includegraphics[width=\textwidth]{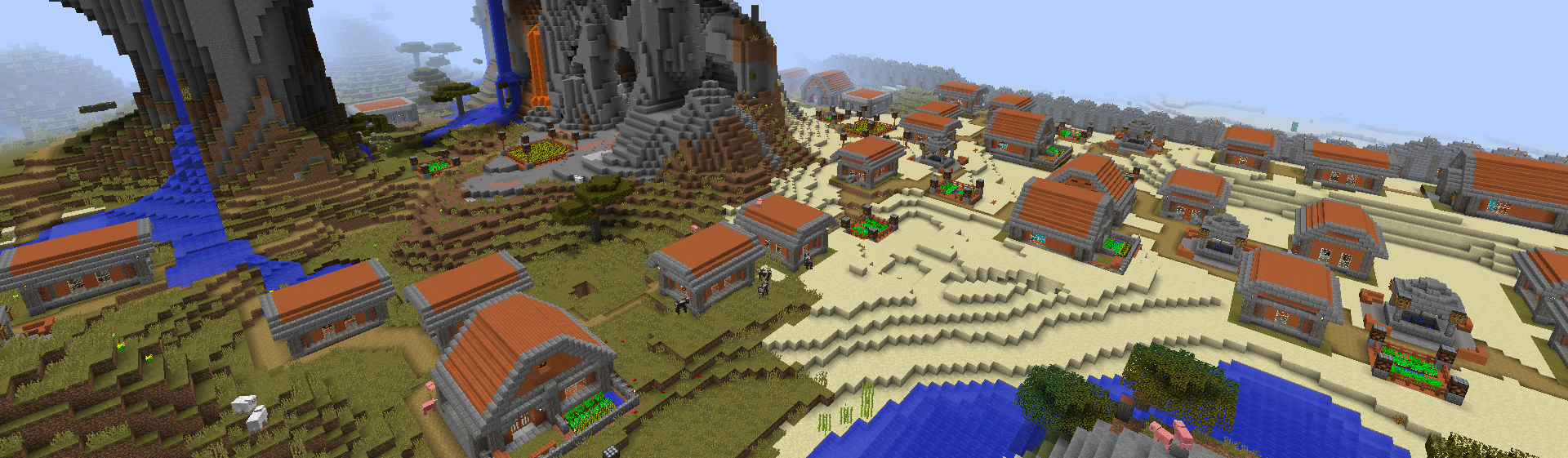}}
\date{}
\author[1]{Christoph Salge\thanks{Corresponding author, christophsalge@gmail.com. All other authors in alphabetical order. Authors are those who provided an experience text, or organizers of the GDMC competition who contributed to editing this article.}}
\author[2]{Claus Aranha}
\author[3]{Adrian Brightmoore}
\author[4]{Sean Butler}
\author[5]{Rodrigo Canaan}
\author[6]{Michael Cook}
\author[5]{Michael Cerny Green}
\author[7]{Hagen Fischer}
\author[8,9,6]{Christian Guckelsberger}
\author[10]{Jupiter Hadley}
\author[1]{Jean-Baptiste Hervé}
\author[11]{Mark R Johnson}
\author[12,13]{Quinn Kybartas}
\author[14]{David Mason}
\author[15]{Mike Preuss}
\author[16]{Tristan Smith}
\author[17]{Ruck Thawonmas}
\author[5]{Julian Togelius}
\affil[1]{University of Hertfordshire, UK}
\affil[2]{University of Tsukuba, Japan}
\affil[3]{theworldfoundry.com}
\affil[4]{University of West of England, UK}
\affil[5]{New York University, US}
\affil[6]{Queen Mary, University of London, UK}
\affil[7]{Pascal Gymnasium M\"unster, Germany}
\affil[8]{Aalto University, Finland}
\affil[9]{Finnish Center for Artificial Intelligence}
\affil[10]{jupiterhadley.com}
\affil[11]{University of Sydney, Australia}
\affil[12]{McGill University, Canada}
\affil[13]{Concordia University, Canada}
\affil[14]{University of Warwick, UK}
\affil[15]{Universiteit Leiden, Netherlands}
\affil[16]{University of Bath, UK}
\affil[17]{Ritsumeikan University, Japan}
\renewcommand\Authands{ and }

\maketitle
\vspace{-1cm}
\begin{abstract}
The GDMC AI settlement generation challenge is a PCG competition about producing an algorithm that can create an ``interesting'' Minecraft settlement for a given map. This paper contains a collection of written experiences with this competition, by participants, judges, organizers and advisors. We asked people to reflect both on the artifacts themselves, and on the competition in general. The aim of this paper is to offer a shareable and edited collection of experiences and qualitative feedback - which seem to contain a lot of insights on PCG and computational creativity, but would otherwise be lost once the output of the competition is reduced to scalar performance values. We reflect upon some organizational issues for AI competitions, and discuss the future of the GDMC competition.  
\vspace{0.5cm}
\end{abstract}

\section{Introduction}

The GDMC AI Settlement Generation Challenge \cite{salge2018generative,salge2020ai} in Minecraft \cite{game:Minecraft} is an annual (since 2018) competition, where participants submit code capable of generating a settlement on a given Minecraft map. The submitted generators are then applied to three different maps, previously unseen by the participants and provided by the organizers. Making generators that can adapt to various types of terrain, and even produce settlements that reflect the peculiarities of said terrain is an essential part of the GDMC challenge. Another key element of the challenge is that there is no computable function that determines the quality of the generated settlement - the algorithm has to create a design appropriate to a given scenario with ill-defined goals \cite[Chapt.~8.]{smith2012mechanizing}. For evaluation the maps are sent out to a range of human judges, including experts from fields such as Minecraft Modding, Game Design, AI and Games research, Architecture, City Planning, and to volunteers who applied out of their own initiative. All judges are asked to look at the three maps for each entry, explore the settlements, and then score each generator from 0 to 10 in four categories - Adaptivity, Functionality, Evocative Narrative and Aesthetics. The judges are given a list of questions that should illustrate the categories (see \cite{salge2018generative}). In short, Adaptivity is about how well the generated settlement changes in reaction to different input maps. Functionality is about both game-play and fictional affordances being provided by the settlement. Evocative Narrative concerns how well the settlement tells a story about the people who supposedly live in it and how it came about. Aesthetics are less about how beautiful the settlement is, and more about avoiding the very simple design errors that are immediately obvious to a human but not to an algorithm. Scores range from 0, for no discernible effort, over 5 for settlements where it becomes unclear if this was done by a human or AI, to 10, for an artifact that looks superhuman, or could only have been built manually by a team of experts with a lot of time. The detailed guide for the judges is also available online for competitors as reference\footnote{For detailed criteria and other information see: \url{http://gendesignmc.engineering.nyu.edu/}}.

We originally had several aims when designing this competition \cite{togelius2014run}. For one, we wanted to increase interest and stimulate work in the field of procedural content generation (PCG)~\cite{short2017procedural,katecompton2016,colton2012computational,shaker2016procedural,liapis2014computational} and computational creativity \cite{colton2012computational}. While we wanted the competition to be accessible to students and the general public, we also wanted it to serve as a test bed to try out and compare different PCG approaches and techniques. In contrast to other ``citizen science'' approaches we were not just interested in the general public working for us, but were genuinely hoping that an open approach to this problem might lead to the development of new ideas that we, in academia and the industry, could learn from. To this end, we tried to design a competition that is not biased towards a certain approach or technique, such as deep learning or genetic algorithms, but rather provides a leveled playing field, as far as this is even possible \cite{canaan2019leveling}.

The lack of a clearly computational quality metric was also a deliberate design feature and motivation of the GDMC competition; as a secondary goal, the present human evaluations might inform the necessary features of such computational measures as a future engineering challenge. 


Many modern PCG methods are based, directly or indirectly, on a notion of optimizing an objective function. While this line of thinking has seen many successes, it also limits what content generation can be to what can at present be quantified. If we based our competition on fulfilling some well-defined objective, we would be liable to the effects of Goodhart's law \cite{goodhart1984problems,strathern1997improving}, according to which a well-defined measure ceases to be useful when it starts being used as a target. In other words, competitors might have created generators that achieved a high value according to an objective function while generating settlements that were unsatisfying to a human. Additionally, creating a meaningful yet well-defined function to optimize for also proved quite hard \cite{canaan2018towards}.

The approach we chose can instead be likened with efforts to create open-ended \cite{stanley2019open,grbic2020evocraft} agent behavior in artificial intelligence. As far as we are aware, all existing open-ended AI research is concerned with the behavior of agents in an environment; this competition is an attempt to bring the open-ended mindset to creating generators of environments.
The GDMC competition also differs from existing PCG work and competitions \cite{khalifa2016general,khalifa2017general,stephenson20182017} in that it focuses on \emph{holistic} and \emph{adaptive} content generation. Holistic PCG means that we are not looking at the generation of one aspect of the environment on its own, but rather at all of them together: buildings, paths, natural features, backstories, and so on. Adaptive PCG means that the generators must work with a complex input that they have been given. In this case, it means that the generators are provided with maps (unseen by the generator designers) and must generate a settlement that works with that map. This requirement was in part introduced to counteract the submission of generators that simply create the same settlement over and over (in the extreme, a ``generator'' could be a single, a priori and manually designed settlement). However, the topic of adaptive PCG is an interesting one in its own right, so we decided to lean into this aspect. 

On the critical side, it is somewhat ironic that after deliberately not reducing the ``creative'' output of the generators to a simple scalar value, we ask the judges to score them on a scale from 1 to 10 which, after calculating the average, is used to determine the winner. 
It is telling that several of our participants were actually not that interested in their scores - but showed much more appreciation of the additional notes and written feedback provided by the judges. In the first year this feedback was only sent directly to participants, but in the later years we also published the feedback given to the participants on our website and on Discord for all involved parties. It became quickly evident that it contained a lot of interesting thoughts, anecdotes, etc. that were deeply insightful for the advancement of both, the GDMC competition in particular, and computational creativity in general.

This paper is an attempt to collect, preserve, summarize and discuss this feedback, and then provide it in a publishable form to those interested. To this end, we have contacted the judges and participants from the previous years and asked them to provide us some form of written impressions about their experience with the competition. We also allowed for submissions from our community, by advertising this project on our social media channels and via the GDMC's community Discord channel. For all those wanting to participate, we provided a list of question prompts (see Appendix~\ref{ref:prompts}), but also expressed our interest in any written text about their experience with the competition. The question prompts were written to elicit responses related to the artifacts (both positive and negative aspects), but also to the competition itself, the way it is judged, and how it relates to computational (co-)creativity. Participants were given the freedom to address any of these points, and answer all, some or none of the questions. We collected all submitted texts, performed minor formatting but no content edits, and now show them in Appendix~\ref{ref:texts}. 

In the remainder of this paper, we provide a general overview of earlier findings on the importance of textual and less structured feedback, and summarize insights from the experience write-ups for this paper specifically. This write-up also relies on discussions we had on various social media channels, which can all be found via our website \footnote{\url{http://gendesignmc.engineering.nyu.edu/}}.

\section{Summary}


\subsubsection{Participation}

The GDMC competition has a growing number of participating teams, with 4, 6 and 11 submissions in 2018, 2019 and 2020. Among our competitors are hobbyists and academic researchers, and in particular university students who have created a generator as part of their coursework. Some of the past approaches that have been used or developed for the GDMC competition have been published as peer-review papers \cite{green2019organic,van2020declarative}. 

\subsubsection{Code Reuse}
Encouraging modular development, facilitating reuse of solutions and lowering the barrier of entry were also inter-related points cited by various respondents, which were both seen as positive and negative. As the competition unfolded over the years, techniques for solving common problems such as calculating heightmaps, removing trees and path-finding needed to be independently implemented and re-implemented by various participants. Our judges reported that the establishment of best practice solutions led to an increase in overall quality, as some things are solved, and work that goes beyond these basics can be attempted. Others criticized the lack of creativity and surprisal they experienced when seeing the same building or solution over and over again - such as the iconic high-rises first introduced by Eduardo Hauck in 2019. As returning participants incorporate more and more of these solutions in their settlements, this could create a perceived barrier of entry for new participants if they feel like these features are needed to compete. While we encourage participants to make their code public after submission, reverse-engineering an existing solution and incorporating it into a new settlement are non-trivial tasks.

We, the organizers, are interested in facilitating positive code reuse by identifying some of these tasks with high reuse potential and providing modules to address them. Examples of how to incorporate these modules, such as a tutorials or complete settlement generators showcasing the modules, would be desirable. It would also be possible to provide incentives for participants and the larger community to contribute to the effort, e.g.~by creating categories for best code documentation or best standalone modules. On a related note, the entry barrier for the Chronicle Generation challenge \cite{Salge2019} is currently particularly high as it requires a team with proficiency both on settlement generation and storytelling. The availability of such modules could particularly benefit teams with skills in either of these domains.

\subsubsection{Large Scale Adaptation}
Another open question is how to delegate more responsibility for high-level decisions to the generator and thus ``climb the computational creativity meta-mountain'' \cite{colton2009seven}. Currently, designers are typically responsible for both, choosing a high-level theme, and translating that theme into elements such as structures, materials, decorations etc. To a large extent, the responsibilities delegated to the generator occur at a lower level, such as selecting suitable spots with appropriate terrain for the structures, laying down paths between various points of interest and diversifying the settlement by combining the elements in different ways.

While delegation of creative responsibilities is an ongoing challenge in the field of computational creativity \cite{colton2012computational} and not exclusive to GDMC, one respondent of the survey highlighted a way in which the judging process discourages participants from investing development time into the delegation of high-level tasks: a generator making a high-level decision runs the risk of not only making a poor choice on a given evaluation map but also of not showcasing a fair share of its high-level capabilities with a limited number of evaluation maps, especially considering the chance of repetition of high-level themes. A problem also often cited to prevent the use of interactive or branching narratives in commercial computer games - something that is great might never be seen by the player. Increasing the number of evaluation maps enough to mitigate this issue would likely be infeasible without placing an undue burden on judges under the current process, but might be feasible under a potential crowd-sourced judging process.

\subsubsection{Qualitative Feedback}

Already in the first year it became very clear how powerful the textual feedback would be in this competition. Themes such as a lack of bridges over water, absence of light, or big, unique set pieces, would pop-up in the qualitative feedback, be discussed in the social channels, and then either reappear or be solved in subsequent years. In many cases the discussion would identify some features of the previous year's winner, and subsequent submissions would aim to incorporate them. Based on the urging of several community members, we now also make the textual feedback accessible to a wider public with potential future participants. The textual feedback was also, in parts, surprising to the organizers, and showed that there are elements in the evaluation we did not previously consider.

\subsubsection{Video Judging}

\begin{figure}[ht]
  \centering
  \includegraphics[width=\linewidth]{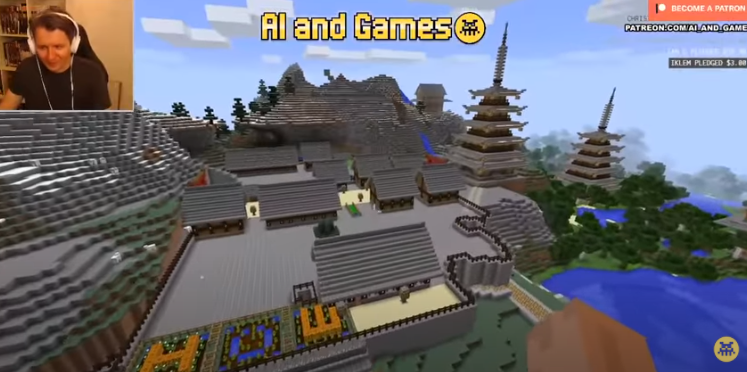}
  \caption{Image from the live-judging stream of Dr. Thompson's channel ``AI in Games'' looking at a settelement generated by the ICE{\_}JIT algorithm.}
  \label{fig:ai_in_games}
\end{figure}

Two of our judges (Jupiter Hadley and Tommy Thompson) used YouTube and similar platforms to broadcast their judging sessions. This found very positive resonance. Many of our later participants reported that this was the pathway they found out about the competition. In particular, the live judging session with Dr. Thompson (Fig.~\ref{fig:ai_in_games}) was well attended online, saw lively discussions, and attendance of several of the competitors. Comments indicate that for many of the participants winning is less important then to showcase their work and receive attention and feedback from professionals. When we discussed potential prizes that would work for our diverse participants one suggestion was the ability to speak to a Game AI professor about their work for 15 minutes and get some advice or feedback. 

Observing the live judging session also provided great insight into the importance of genuinely interacting with the created artifact. All our judges are encouraged to actually walk through the settlements, but we, of course, cannot check that. Seeing someone actually interact with the generated artifacts, and exclaim in excitement, was illustrative in figuring out what parts or elements evoke emotions in people. This, and playing the settlements itself, also gives an opportunity to experience some of the more ephemeral effects of interacting with PCG. As one commenter pointed out - there is a slightly different relationship with PCG artifacts compared to human-made artifacts when it comes to ownership: changing someone else's Minecraft settlement without their permission is considered rude, while doing so to an AI is less so. This was also evidenced by Dr. Thompson in his live judging session - as he apologized for smashing apart houses to then realize that it probably does not matter. This raises the interesting question whether this perceived need for protection, which we usually assign to pieces of art and creative works by humans, could give us a yardstick to measure the quality and human-likeness of PCG artifacts. 

Finally, the video judging also demonstrated through its comments how the cultural and biographical backgrounds of a person matter for their relationship to the PCG artifacts. This is not surprising itself - but can be seen as positive in the sense that the complexity of the artifacts designed here has risen to a level were it starts to matter. This was also addressed in ICE{\_}JIT's submissions text, which explicitly discusses the Japanese influences used for their generator. 

\subsubsection{Evaluation}

We specifically queried people about our evaluation methodology, which stands apart from what is used by most existing AI competitions. Not only do we not use an objective evaluation function, we even forgo the use of well established human-centric methods, such as ranking  \cite{Yannakakis2015}, to establish the ``best'' solution. This is, in part, due to the fact that it is unclear if there really is a best generator, or if several interesting and great solutions can stand side-by-side. Several GDMC community members have suggested that we should embrace this concept more fully, and forgo having winners all-together, and rather go for a ``festival of ideas'' approach, where we only give qualitative feedback and celebrate different and interesting ideas. Other suggestions are to introduce achievement tiers, where scores are still given, but the aim is to get to a certain level, rather than to beat others. Finally, there are repeated suggestions to crowd-source the evaluation. We do already publish the generated settlement maps and most generators, and have in the last year even started to host a public server that contains a composite map showing all submitted generators. However, this is only for a post-judging evaluation by the public. Setting up a crowd based evaluation would, if successful, provide us with a statistically more sound evaluation of the settlement, but would probably, against our goals, yield a more results- and less feedback-focused evaluation.

The evaluation debate also brought up some other issues, namely that it is unclear what exactly the ideal generated settlement is: should it be similar to what humans would actually build in a Minecraft game - which can often by quite chaotic and aesthetically jarring - or should it resemble a well designed piece as they are often done by professional speed builders? These issues have also been raised in regard to functionality as a criterion, where the actual functionality of a settlement, such as protection from danger, keeping mobs from spawning, navigability for a human and function blocks, is not the same as having a functionally looking city, that has car lanes or canals in a world without cars or ships. Several of our criteria are just about an imagined functionality, such as a trading harbor, even though there is not need for this in game mechanic terms. 

\subsubsection{Future Direction}

Apart from revising the evaluation criteria, there is also the desire to expand the scope of the competition. We already introduced a bonus competition last year, the Chronicle Challenge \cite{Salge2019}, which asks participants to also produce a book in the Minecraft settlement that tells a meaningful story about the settlement, hence addressing challenges in computational storytelling \cite{gervas2009computational}. There are also requests to permit larger settlements - which would allow for differently scaled building styles. We also discussed several additional forms of adaptivity - such as having a generator continue to build a small settlement started by a human, or to have one settlement generator applied after another. In 2021 we will experiment with both having larger evaluation maps, and maps that have smaller hand-made settlements already present. 

There have also been discussion on moving further towards embodied, co-creative aspects \cite{liapis2016mixed} - by having actual agents that build settlements in the game, or that could built together with humans. There has not been much development on this front, as we believe this would not only be technically challenging for us to set up, but also for participants to get into. Some initial steps could be taken by cooperating or integrating with other existing Minecraft competitions. For example, there is a reinforcement learning focused competition for Minecraft Bots \cite{milani2020minerl,guss2021minerl}, and it might be interesting to see if they could operate in procedural generated settlements. Similarly, there is now a framework for using evolutionary algorithms to build machines in Minecraft \cite{grbic2020evocraft}, and it might be worthwhile to see if that framework could be used for our purposes. We are also currently looking at either using additional or new frameworks, beyond the MCEdit \cite{mcedit} and Java Mod based-approaches that we embrace at the moment. In 2021 we will be using a new framework developed by on of our community members, that allows for interaction with a live Minecraft world via a http interface. This will allows competitors to write clients in a range of programming languages, and also would allow for editing the world live, allowing for a player to observe the generation process as it unfolds. While this is not a evaluation form planned for the coming year, it is something that has been requested several times, as it would allow for the generation of an experience similar to watching the popular time-lapse videos of teams building Minecraft settlements. 

Furthermore, using this http client would also allow us to update to the current Minecraft version. There is also the hope that building a PCG generator in a framework that could be directly modded into the game might make it see more use once developed.  

\subsection{Conclusion}
Overall, we received a lot of positive and constructive feedback - both from participants and the wider academic community. We are planning to keep developing the GDMC competition, and will organize and present the 2021 round of the competition at the Foundations of Digital Games conference. 

\section*{Acknowledgements}
Acknowledgments to the number of unnamed participants and GDMC community members, whose hard work in making the generators discussed here. CG is funded by the Academy of Finland Flagship programme \emph{Finnish Center for Artificial Intelligence} (FCAI). RC gratefully acknowledges the financial support from Honda Research Institute Europe (HRI-EU).

\bibliographystyle{acm}
\bibliography{sample-base}

\begin{thebibliography}{10}

\bibitem{canaan2018towards}
{\sc Canaan, R., Menzel, S., Togelius, J., and Nealen, A.}
\newblock Towards game-based metrics for computational co-creativity.
\newblock In {\em 2018 IEEE conference on Computational Intelligence and Games
  (CIG)\/} (2018), IEEE, pp.~1--8.

\bibitem{canaan2019leveling}
{\sc Canaan, R., Salge, C., Togelius, J., and Nealen, A.}
\newblock Leveling the playing field: Fairness in ai versus human game
  benchmarks.
\newblock In {\em Proceedings of the 14th International Conference on the
  Foundations of Digital Games\/} (2019), pp.~1--8.

\bibitem{colton2009seven}
{\sc Colton, S.}
\newblock Seven catchy phrases for computational creativity research.
\newblock In {\em Dagstuhl Seminar Proceedings\/} (2009), Schloss
  Dagstuhl-Leibniz-Zentrum f{\"u}r Informatik.

\bibitem{colton2012computational}
{\sc Colton, S., Wiggins, G.~A., et~al.}
\newblock Computational creativity: The final frontier?
\newblock In {\em ECAI\/} (2012), vol.~12, pp.~21--26.

\bibitem{katecompton2016}
{\sc Compton, K.}
\newblock So you want to build a generator….
\newblock
  \url{http://galaxykate0.tumblr.com/post/139774965871/so-you-want-to-build-a-generator}.
\newblock Accessed Feb 2016.

\bibitem{gervas2009computational}
{\sc Gerv{\'a}s, P.}
\newblock Computational approaches to storytelling and creativity.
\newblock {\em AI Magazine 30}, 3 (2009), 49--49.

\bibitem{goodhart1984problems}
{\sc Goodhart, C.~A.}
\newblock Problems of monetary management: the uk experience.
\newblock In {\em Monetary theory and practice}. Springer, 1984, pp.~91--121.

\bibitem{grbic2020evocraft}
{\sc Grbic, D., Palm, R.~B., Najarro, E., Glanois, C., and Risi, S.}
\newblock Evocraft: A new challenge for open-endedness.
\newblock {\em arXiv preprint arXiv:2012.04751\/} (2020).

\bibitem{green2019organic}
{\sc Green, M.~C., Salge, C., and Togelius, J.}
\newblock Organic building generation in minecraft.
\newblock In {\em PCG Workshop at the Foundations of Digital Games
  Conference\/} (2019).

\bibitem{guss2021minerl}
{\sc Guss, W.~H., Codel, C., Hofmann, K., Houghton, B., Kuno, N., Milani, S.,
  Mohanty, S., Liebana, D.~P., Salakhutdinov, R., Topin, N., Veloso, M., and
  Wang, P.}
\newblock The minerl 2019 competition on sample efficient reinforcement
  learning using human priors, 2019.

\bibitem{hingston2009turing}
{\sc Hingston, P.}
\newblock A turing test for computer game bots.
\newblock {\em IEEE Transactions on Computational Intelligence and AI in Games
  1}, 3 (2009), 169--186.

\bibitem{karpov2013believable}
{\sc Karpov, I.~V., Schrum, J., and Miikkulainen, R.}
\newblock Believable bot navigation via playback of human traces.
\newblock In {\em Believable Bots}. Springer, 2013, pp.~151--170.

\bibitem{khalifa2017general}
{\sc Khalifa, A., Green, M.~C., Perez-Liebana, D., and Togelius, J.}
\newblock General video game rule generation.
\newblock In {\em Computational Intelligence and Games (CIG), 2017 IEEE
  Conference on\/} (2017), IEEE, pp.~170--177.

\bibitem{khalifa2016general}
{\sc Khalifa, A., Perez-Liebana, D., Lucas, S.~M., and Togelius, J.}
\newblock General video game level generation.
\newblock In {\em Proceedings of the Genetic and Evolutionary Computation
  Conference 2016\/} (2016), ACM, pp.~253--259.

\bibitem{liapis2016mixed}
{\sc Liapis, A., Smith, G., and Shaker, N.}
\newblock Mixed-initiative content creation.
\newblock In {\em Procedural content generation in games}. Springer, 2016,
  pp.~195--214.

\bibitem{liapis2014computational}
{\sc Liapis, A., Yannakakis, G.~N., and Togelius, J.}
\newblock Computational game creativity.
\newblock In {\em ICCC\/} (2014), pp.~46--53.

\bibitem{milani2020minerl}
{\sc Milani, S., Topin, N., Houghton, B., Guss, W.~H., Mohanty, S.~P., Vinyals,
  O., and Kuno, N.~S.}
\newblock The minerl competition on sample-efficient reinforcement learning
  using human priors: A retrospective.
\newblock {\em arXiv preprint arXiv:2003.05012\/} (2020).

\bibitem{game:Minecraft}
{\sc Persson, M.}
\newblock {M}inecraft, 2011.

\bibitem{salge2020ai}
{\sc Salge, C., Green, M.~C., Canaan, R., Skwarski, F., Fritsch, R.,
  Brightmoore, A., Ye, S., Cao, C., and Togelius, J.}
\newblock The ai settlement generation challenge in minecraft.
\newblock {\em KI-K{\"u}nstliche Intelligenz 34}, 1 (2020), 19--31.

\bibitem{salge2018generative}
{\sc Salge, C., Green, M.~C., Canaan, R., and Togelius, J.}
\newblock Generative design in minecraft (gdmc) settlement generation
  competition.
\newblock In {\em Proceedings of the 13th International Conference on the
  Foundations of Digital Games\/} (2018), pp.~1--10.

\bibitem{Salge2019}
{\sc Salge, C., Guckelsberger, C., Green, M.~C., Canaan, R., and Togelius, J.}
\newblock Generative design in minecraft: Chronicle challenge.
\newblock In {\em Proceedings of the 10th International Conference on
  Computational Creativity\/} (2019).

\bibitem{shaker2016procedural}
{\sc Shaker, N., Togelius, J., and Nelson, M.~J.}
\newblock {\em Procedural content generation in games}.
\newblock Springer, 2016.

\bibitem{short2017procedural}
{\sc Short, T.~X., and Adams, T.}
\newblock {\em Procedural Generation in Game Design}.
\newblock CRC Press, 2017.

\bibitem{smith2012mechanizing}
{\sc Smith, A.~M.}
\newblock {\em Mechanizing exploratory game design}.
\newblock PhD thesis, UC Santa Cruz, 2012.

\bibitem{stanley2019open}
{\sc Stanley, K.~O.}
\newblock Why open-endedness matters.
\newblock {\em Artificial life 25}, 3 (2019), 232--235.

\bibitem{stephenson20182017}
{\sc Stephenson, M. J.~B., Renz, J., Ge, X., Ferreira, L.~N., Togelius, J., and
  Zhang, P.}
\newblock The 2017 aibirds level generation competition.
\newblock {\em IEEE Transactions on Games\/} (2018).

\bibitem{strathern1997improving}
{\sc Strathern, M.}
\newblock ‘improving ratings’: audit in the british university system.
\newblock {\em European review 5}, 3 (1997), 305--321.

\bibitem{togelius2014run}
{\sc Togelius, J.}
\newblock How to run a successful game-based ai competition.
\newblock {\em IEEE Transactions on Computational Intelligence and AI in Games
  8}, 1 (2014), 95--100.

\bibitem{van2020declarative}
{\sc van Aanholt, L., and Bidarra, R.}
\newblock Declarative procedural generation of architecture with semantic
  architectural profiles.
\newblock In {\em 2020 IEEE Conference on Games (CoG)\/} (2020), IEEE,
  pp.~351--358.

\bibitem{mcedit}
{\sc Vierra, D.}
\newblock Mcedit.
\newblock \url{https://github.com/mcedit/mcedit}, 2015.

\bibitem{Yannakakis2015}
{\sc Yannakakis, G.~N., and Martínez, H.~P.}
\newblock Ratings are overrated!
\newblock {\em Frontiers in ICT 2\/} (2015), 13.

\end{thebibliography}

\appendix

\section{Question Prompts}
\label{ref:prompts}

The participants who submitted texts to this anthology were given a set of question prompts to give a general idea of what this collection was about. They were also told that they could write about whatever they chose that they felt was meaningful to share in terms of their experiences with the competition. Following this block are the numbered questions prompts. Several participants submitted text with inline answers, and for brevity we removed the original question prompt and replaced it with the numbers.

\begin{enumerate}
    
\item Looking at the generator output from 2020, or even earlier years, was there something you found particularly good? Something where you were surprised that it was made by an AI / algorithm?

\item Where do you think the current challenges lie / what do you think are common shortcomings that you think need to be overcome, or should be overcome? Is there something in the generators where you think that an algorithm will never be able to do this?

\item Do you think the way settlements are evaluated by humans currently works? Do the categories and guiding questions capture the overall quality of the settlements, or is there something else that is good or bad about them that you find not reflected in the judgement?

\item Is there something you noted – or something you felt that is worth pointing out, while you looked at the settlements? Any anecdotes worth sharing?

\item If you are a person who built one of the generators, what were the challenges, what did you find easy? Is there something that surprised you while you built the generator? Some good advice you want to give for others doing this.

\item Keep in mind that the overall goal is not just to make the best Minecraft settlement generator, but to get a better understanding on how digital creativity works, particularly those that must deal with existing content. Does the GDMC competition do a good job with gaining insights into this? Is there something you learned or something that was developed that you think will help with computational creativity in general? Is there something we should change about the GDMC competition to move it more towards this ideal? Feel free here to muse about any connection between computational creativity in general, and the GDMC in particular.

\item Also, keep in mind that we are planning, over time, to move more towards interactive human-AI co-creativity – so maybe at one point we want to have an AI that can build together with a human, maybe even in real time. Do you think this is a good goal? What do you think the challenges in this framework would be for this? Would this still work?
\end{enumerate}

\section{Submitted Texts}
\label{ref:texts}

\subsection{Michael Cook, Advisory Board Member and Judge}

1: One of the more significant leaps forward has been the addition of small narrative details that convey a sense of place or history. This isn't present in earlier entries, and generally isn't something we see in a lot of classic content generators. However these still feel more on the side of human-authored than AI-mediated, it's clear that the narrative here has been written in advance by a human.

In terms of features that have more AI involvement in their placement, I think a recurring point that really impresses me is when a generator uses space to position an important building and the AI then integrates it into the result. A good example of this is the placement of watchtowers at high points in the map, with pathing algorithms clearing a way to get there. This really captures my imagination, following the path and thinking about people ascending to this isolated guard post. It's still got the hand of a human designer in its inspiration, but the placement and the way in which terrain is dealt with is handled by the AI, so it feels like a pleasing blend of approaches. 

2: A major challenge is understanding environmental context. The generators fail at this in small ways (for example, mistaking a very tiny puddle for a lake and putting a huge bridge over it). But they also fail in more serious ways, on a larger scale. Every generator broadly tries to build the same settlement regardless of the space, and ones that don't (for example, one generator this year failed on a tiny island map and only placed one or two buildings) mostly are failure cases. 

The difficulty with settlement generation, looking ahead to the far future, is similar to that of narrative generation or high-concept content generation like planets in No Man's Sky. Most PCG systems work at the lowest-level of detail, and the human designer influences at the highest level of detail. For example, the human making the settlement generator might decide to place a watchtower at the highest point in the map (a high-level design decision) and the AI has to find a path of blocks that connect the town to the watchtower (a low-level problem solved by pathfinding algorithms and other techniques). Invention of high-level concepts is challenging because it asks the settlement generator to make high-level design decisions, and these typically require a lot more knowledge, especially knowledge outside of the domain. A good Minecraft settlement might be inspired by movies, other games, real-life locations or even creator emotions. I don't think it's impossible for an AI to do this, but I think for a long time we'll struggle to make progress in these areas.

3: I think human evaluation is the best we can do, and I think it suits the nature of the competition - it helps entrants get good feedback, it makes the process interesting and adds spice to announcing the results. I think in terms of strict methodology, it's not the best way to truly gauge the quality of a settlement generator, but I think that's a really challenging research question in itself. 

I have discussed this in the past with the organisers, but I personally would probably lean entirely into aesthetics if I were running the competition, and not have categories for mechanics (such as lighting to keep away monsters, or resource generation). I can understand including these though, they're just not aspects that interest me personally, and as someone who doesn't play Minecraft regularly I often forget some of these factors during evaluation. 

Overall, while human judging isn't perfect, I think it's a consistent and effective way of assessing settlements. Much like using numeric scores for videogames or films, it's not an exact science, and scores aren't necessarily perfectly comparable across entries. But you do eventually end up getting a feel for what a "good" system looks like, or an "average" system. My one criticism is probably the nature of the scales themselves. I don't know how other judges score, but I think the standard of "a human could have produced this" is potentially misleading as it's hard to distinguish where the generator expended effort, what decisions it made, and so on. I often find myself wanting more space in the lower end of the scoring scale, because I'm not going to use the 5-10 section for a very long time, but there are interesting grades between 1,2 and 3 that I often want to explore.

I think ultimately I might prefer to simply rank each year's entries relative to one another rather than score them on an objective scale, if that makes sense. But these are fairly small complaints and the current system works fine. As long as we don't get news stories written about "AI is officially better than Frank Lloyd Wright!!!"

4: I think discovering small details for the first time is always delightful. Also I spend a lot of time flying around, but walking has really improved my perception of the space. I think my best experiences have come when a settlement takes advantage of dramatic landscape, and you get to explore it at eye level and see houses appear and disappear around hillsides, or climb to the peak of a mountain. I'd like to see more settlements really emphasise the design of spaces and what it feels like to travel around.

6: In my opinion the GDMC competition is a very important contribution to the study of digital creativity. There are not enough AI challenges that emphasise open-ended creativity like this, or that set such large and complex goals. I don't really care if an AI can play Go or Chess better than a human, but I am really interested in what we can get AI to do in creative spaces, where the emphasis is less on being superhuman and more on being interesting or different. I'd like to see GDMC double down on this emphasis. One thing I would really love is a separate track for photographer bots. Given a settlement, the bot goes around and takes 3-5 photographs and submits the set. Judges select the best photographer. I think GDMC can use this creative space it's carved out to set more creative challenges, like the chronicle challenge, that encourage researchers to do different things with AI. It's a great initiative and I hope to see it keep growing.

7: I think co-creativity is a good goal to work towards for GDMC, as a complementary goal to autonomous settlement design. I think this potentially is more challenging to evaluate, and to design. I could imagine AI being given starter settlements to complete, for example, with judging categories added for things like how well the AI fits in with the original style, or to what degree they try to innovate or converge on a style. 

It could also have interesting overlap with the chronicle challenge - can the AI write a chronicle for a human designed settlement, can it design a settlement based on a human chronicle. Lots of cool opportunities for overlap between human and AI creativity.

\subsection{Mark R. Johnson, Advisory Board Member and Judge}

1) What probably stands out for me is the variety in the buildings, especially the settlements that clearly had a number of different building archetypes and distributed them in larger or smaller numbers depending on what the (implied) nature of the building was. That struck me as a very compelling and very convincing bit of generation which takes account of the larger picture as well as the individual components.

2)I think a main area for GDMC submissions to improve is in adapting to the terrain given, rather than changing the terrain to meet the requirements of a less flexible generator. A few of the submissions in 2019 did this well and I'm sure those in 2020 did so too; but there's still a long way to go. This is both something which requires very "small-scale" adjustments in the generator - such as taking account of a block up or down in terrain - but also major changes, such as finding ways to work around or incorporate hills, mountains, rivers, canyons, or whatever. I'm particularly excited to see where this particular aspect goes in the future - and to move away from the generators that just flatten out a huge grid of terrain and then build a town on top.

3) I think the current evaluation metrics honestly capture the main things we should be looking for here. Were we not asking people to generate a *settlement* different metrics would be appropriate of course, but for a settlement - and everything that comes with that in terms of the use of space, how time and progression are visible in such a place, and any cultural or social implications - I think the current reviewing method is ideal. Of course, if future competitions yield even more sophisticated entries we might need to develop them further.

4) I remember a couple of buildings that spawned with wild animals in them, which gave me a bit of a chuckle. Managing inside and outside space is something that might be interesting to keep an eye on going forward. Equally, as someone who doesn't play or engage with Minecraft particularly beyond this competition, it's important to keep in mind the differences and overlaps between the game of Minecraft and the outputs of the generator. I think there's a complex relationship here between the game, its affordances and assumptions, and what a generator within that game can produce, which we should think about more. 

5) N/A

6) In my view: yes, absolutely. I think it's a very valuable competition, and more broadly I also think this has got to be a good method for simply getting people interested in the domain, whether people considering pursuing research or further study, or just hobbyists who are intrigued by the area. Minecraft is so universal and so accessible that the GDMC competition strikes me as potentially a fantastic starting-point for lots of people. (This is another reason I'm keen for us to push advertising of the competition much more widely, continue trying to get some kind of a mention in MineCon, etc). 

7) I think this would be really interesting - not more so than the present challenge, but certainly equally so. It would add some extra dimensions to what we'd be judging them on, i.e. we'd also want to judge the AI on how usable it was, for instance, or the user interface if relevant, these kinds of things, or how well the AI responds to what we want it to do, how well it offers us suggestions, etc. In a sense we would be marking on "process" as well as "outcome", whereas in the competition at present we are basically only marking on "outcome". (Though for what it's worth, my inclination would be to focus on and grow the existing competition further, at least for a few years - I think this is a competition with a ton of untapped potential at the moment even before we get into any "collaborating with AI" stuff). 

\subsection{Tristan Smith, Judge}

1: The process of evaluating the generators each year is a joyful experience, as it centres on surprise and discovery --- each entry has its own approach and effect on the landscape, and across three maps it is possible to start to unpick an understanding of what it does and does not do; or only does by chance, or occasionally tries to do but miscalculated. Thorough assessment requires exploration, investigation, and identification of techniques, commonalities and exceptions. For example one of the recurring methods visibly uses some sort of Voronoi approach to parcel up the land, and it is instructive to see where that succeeds or runs into issues --- and how individual entries have inherited approaches and ideas, or introduced something entirely new to the pool.

As is expressed in more detail later, some of the most interesting experiences while judging in 2020 were both those that were inhuman, and those that were very human. A standout moment was while judging the first of the settlements produced by David Mason’s generator; turning around and first seeing the huge central open pit mine. In comparison to the alterations produced by generators previously this was an unexpected change on a jaw-dropping scale. Several of the other generators in 2020 made similarly massive additions, and in each case these became a consistent key signature of that generator, on a larger-than-human scale. In contrast, The World Foundry’s entry with named and visually-differentiated builders provided a distinctly human experience: the sensation of being able to recognise another player’s work because of their distinctive style. Whereas other settlements’ buildings are frequently identifiably produced by that particular generator (or that generator’s tower block/modular building/anvil tower subsystem), The World Foundry’s system instead cemented an association between a clear visual signature and in-fiction NPC names. This mirrors the experience of learning human players' variations in style: it is possible to play on a server with the same people consistently and begin to identify personal approaches to location, form and/or materials --- and it felt like gaining those same connections. It’s tough to express some of these concepts in the assigned scores for generators or the notes that accompany them, and so the deliberate opportunity to reflect here has also been valuable.
 
2: It has been interesting to see the progression in generator capabilities year-on-year, the recurrence of popular techniques, and even clear code lineages shared between entries. One of the ongoing challenges is that while each of the entrants will be facing many of the same problems as each other, and as other entrants in previous years, the diversity of approaches means that it will not always be easy to extract a previous generator’s solution to a particular problem and re-use it --- there are no fundamental building blocks for solving common issues. Though this helps to individuate entries based on which challenges they attempt to tackle and the approaches they use to do so, it also potentially raises the perceived barrier for entry, as new generators must do more to seem competitive.

One example: an issue in some earlier years’ settlements was the incomplete removal of trees and/or grass, resulting in floating leaves and hard edges in vegetation that left clear evidence of where the generator had made alterations. Though this is mostly cosmetic and only largely impacted the aesthetics of the settlement, occasionally it meant that doors or pathways were blocked by leftover leaf blocks. Since then the standard has raised and this is rarely an issue --- but that doesn’t yet mean that it is a solved problem, as new generators will still need to re-solve it. Similar common issues include making paths traversable across height differences of more than one block, laying out farms for proper irrigation and ensuring they won’t freeze, and constructing reasonable foundations for structures on cliffs or hillsides. As the competition continues, the community may yet develop standard solutions to some of these problem.

Some of the more complex outstanding challenges relate to embodiment and the ease of `living in’ the generated settlements, and these would represent a significant advance on existing generators if solved. A major category of these consist of ensuring that it is easy and efficient to travel between important locations, rather than `merely’ possible (which is the level that most generators are still operating at) --- this includes making paths direct, but also levelling out short deviations in elevation, constructing hypotenuse stairways where possible instead of path + ladders, ensuring that the only entry to a farm or building faces a reasonable direction of approach, and bridges. An algorithm for accurate connection between two raised locations using a bridge, whether over an existing pathway or simply to avoid the need to climb down and then up, would be a strong improvement on the current state of generators' contextually-aware adaptations.

Another embodiment-related concern is that of safety: one of the clearest differences between a player-inhabited area and many of the generated settlements is an attention to prospective sources of danger. Attempting to evaluate an entry only to find that portions of it are already burning down is not an uncommon occurrence, and while some teams this year took explicit steps to neutralise lava in the settlement area and make it `safe’ this is still far from universal. To a lesser degree, there are still occasional instances of flooding where a generator makes unsafe changes near to water, or of collapse where changes are made near to (or consisting of) unsupported sand blocks. A further common issue is sparse lighting coverage --- though several entries ensure that their building interiors are well-lit, fewer take care to light exterior entries with the same thoroughness as cautious human players and after a couple of daylight cycles most maps are still teeming with mobs. Accurately identifying and accounting for `dangerous’ (dark or fire-hazardous) areas within the generated settlement is clearly an ongoing unsolved problem, and one for which a solution would be necessary if entrants were explicitly seeking to mimic human-constructed settlements.

Though some of these considerations are individually called out within the `Functionality’ category of the evaluation advice (which is in turn based on ``what [the competition organisers] believe the challenges in moving towards more human-like settlement generation are’’), one of the strengths of the current evaluation approach is arguably that none of the criteria are `does this look like a human made it?’. This leaves room for interesting generators that make settlements that humans either wouldn’t or couldn’t (feasibly) produce. Though the winning entry in 2020 was debatably the one that produced settlements whose buildings looked most similar inside-and-out to Minecraft’s own, it also produced vast city walls far larger than many casual human builders would ever construct --- and this was an interesting new feature that hadn’t really been seen before. Similarly, David Mason's entry presented a huge open pit mine on a scale that would make no real sense for humans to attempt, and the Architectural Profiles generator emitted vast alien megastructures unlike anything humans might normally set out to create. And though these three entries varied in how `settlement-like’ they were, and this was reflected in the scores --- they were each separately interesting to explore and appreciate in their own rights. The un-human aspects of their outputs aren’t therefore necessarily flaws as such, but affordances of a system that doesn’t tire and to which scale is almost irrelevant, and which therefore explores regions of the space of settlement-like-things unavailable to all but the largest and most-dedicated teams of humans.

Broadly, the current evaluation approach appears to function well with regards to rewarding progress on some of the challenges involved in producing human-like settlements, without overly constraining things by actually requiring that the generated settlements be human-like. 

3: This means that for individual entrants there is an important question: whether to attempt to generate a settlement that is believable within the context of the existing game and its own generation style (i.e. Troy, Ryan \& Trent’s walled village); one that builds on the resources available to produce outputs that more closely echo specific real-world aesthetics (the soaring pagodas and torii gates of ICE{\_}JIT’s entry); or something entirely alien and new (Architectural Profile’s vast blocks and stairways). Each of these potential avenues are interesting and contain their own challenges and experiences, and it seems valuable that the evaluation criteria makes space for all of them. In some ways, this high-level intent choice mirrors the similar choices made by Minecraft mod developers, where the term `Vanilla+’ is used for mods that intend to remain within the `spirit’ of the original game --- which is a useful and informative design direction, but far from the only one that is possible or valued. Likewise, though functional Minecraft-like settlements built on a human scale will clearly be the most familiar to human evaluators, and deserve to continue to score well if executed well, the competition as a whole benefits from diverse generators that do \textit{not} do that, and instead aim to e.g. use interesting or improbable variations on building materials in service of expressing unique styles (The World Foundry), or construct grand aesthetically-pleasing `diorama’ settlements (ICE{\_}JIT).

The current evaluation method allows for this healthy diversity and variety \textit{between} generators, but shapes the priorities of the entrants in ways that perhaps unintentionally discourage variety \textit{within} generators. As each generator is judged on a total of three outputs, it becomes inefficient for entrants to implement too great a variety or context-specialisation in possible productions --- whether these are biome-dependent specialisations for which the right biome might never be selected, or alternate versions of the entry’s own grand concepts like David Mason’s open pit or University of Tsukuba’s monorail. If there are even only two possible settlement centrepieces that the generator might choose from, there is still a 1-in-8 chance that one of the two might never be seen by judges, and the more possibilities there are the greater the likelihood that any development time spent implementing them is effectively wasted, if they do not show up in one of the judged maps. This affects the value of implementing several classes of features that might make for good additions to `a settlement generator’, but not worthwhile inclusions to `an entry to this competition’. In terms of the `10,000 bowls of oatmeal’ metaphor coined by Kate Compton, the reward landscape currently heavily incentivises making a consistently good bowl of (each generators’ version of) oatmeal, and somewhat discourages investing too much time in anything that makes each bowl special, or unique, or surprising. Each generator individually is special, and unique, and surprising, but evaluating on three outputs impedes insight into the variance an individual generator can achieve, and so it’s not necessarily rewarding for entrants to invest on improving on that axis.

Though there is inertial value in preserving the current evaluation approach unchanged in order to facilitate comparison with the entries of previous years, there is also potential for the overall winner to be decided by a combination of scores: from the existing expert evaluation as-is, and additionally either a `counterpoint’ bonus score phase, a crowdsourced `audience participation’ score, or both. A limitation of the need for expert evaluation of the generator outputs is that --- for fairness --- each human evaluator works from the same corpus of sampled outputs, and there is finite time available for judging: hence only three outputs per generator. One possibility for ameliorating the effects of the small sample size on generators’ internal variety would be to offer entrants a `counterpoint’ opportunity, once judging has begun and the selected test maps are publicly known. Generator authors would have until judging ends to select and respond with a location in a fourth known seed, for which their submitted entry produces either an exemplar of what it does best or an output that is most unlike what the judges have seen so far. There could then immediately be a short second judging round in which evaluators can award (or deduct) bonuses to their scores in response. Alternatively (or also), conference attendees or the general public could be invited to form briefer impressions of a wider selection of outputs from each generator, and provide ranked comparative votes to be used in awarding bonuses that would be informed by the wider exploration possible by this cohort. Clearly any changes to the evaluation --- whether these ones or others --- would increase the complexity of organising and running the competition, and so there is a trade-off to be considered between that and the value of reshaping the incentives to reward features and specialisations that might otherwise never be seen.

Intermediate steps could also be possible without too-greatly increasing the burden on volunteer evaluators or competition organisers, such as by adding variants. Each generator is provided with a selection within an existing map, and individual generators have different interpretations of whether the available selection is intended to be the full extent of the settlement, or whether the settlement should just be somewhere appropriate within that bound. Not all generators cope equally well with unexpected biomes or sizes of settlement location selection, as shown in 2020 where a small island was one of the generation locations. A step towards assessing generator variability would be to add a fourth sample, a smaller selection at one of the three previous locations, to see how this affected the settlement produced there without needing to assess the output on a new location from scratch.

4: As noted by others during FDG 2020 there is an altered sense of acceptability when interacting with generated content. An interesting side-effect of exploring swathes of generated settlements is a change in attitude towards ownership --- while the generator did \textit{produce} these artefacts, it is no longer `present’ in the way a player might be, and certainly doesn’t care about what happens next. Though it is impolite-bordering-on-hostile to tamper with another player’s constructions in a normal setting, it feels less taboo to ‘adopt’ generated content – whether that means cautiously poking at a ceiling of unsupported sand blocks to watch them all collapse, or racing to put out a building whose foundations breached lava. This can help to engender a sense of playful discovery --- there are no real risks and no-one whose careful work you’re undoing, just an interesting new environment to explore and to try to understand. This is part of what makes evaluating the generators each year a joyful experience.

\subsection{Jupiter Hadley, Judge}

1: There have been a lot of impressive maps entered into the contests. I am
always impressed when moving buildings are generated - especially the
moving windmills that appeared in one of the previous entries. They
worked so well and made a big impact, and the entire map fit well into
the actual setting around it.

2: I think the biggest challenge seems to be how the buildings blend with
the land. Sometimes, maps look like they have been carved all around
buildings, forcing them into the map. I feel that the flow of the
buildings with the world around them are important, and I do think that
algorithms are able to overcome this, it's just a challenge for
developers to take all of the biomes into consideration and put in the
extra effort to make sure that the buildings do flow with them.

3: I think that experience could be taken into consideration when it comes
to these maps. I feel that there isn't much need in the maps fitting
into small islands, like in the last competition, but that's mainly
because most of the buildings were too big to look good on the islands
and not many of them took the island into consideration.

4: There are a lot of big, sky scrapers entered into these competitions. We
see several of them each year. I feel that these must be less
challenging to make, and are something that often looks the same due to
no internal decorations. I do wish that there was a bit more creativity
when it comes to sky scrapers and more creativity when it comes to the
environment around the individual buildings.

Also, nobody seems to consider lava! It's everyone's worst enemy, as it
will burn up builds.

6: Minecraft is a game most people are familiar with, so adding in
generation - as a player - I can easily see what they have made,
understand what they are going for, and I know the limits of the
creative material within the game. I find that the GDMC does a great job
with showcasing exactly what is generated by the developer and what is
the pre-existing world.

I think the overall competition will help with computational creativity,
as it showcases beautiful worlds, generated, in a well known game.

7: I think interacting with AI is always a good goal, so this sounds very
interesting! I think that the user control over the AI is something to
consider, and the AI's own role in the competition, but it would be very
interesting to see.

\subsection{Quinn Kybartas, Judge}

1: Perhaps I have been working with PCG for too long, but I was generally able to understand, or at least guess, at the algorithms guiding most of the generated settlements I was evaluating. This did not detract from the impressiveness of each entry, or the work/effort involved. Doing even simple generation is extremely challenging once you get into the coding process. Almost all entries have at least had some impressive features to them, such as nice visuals, well laid paths, dense construction, etc. Personally, I would always find myself impressed by any entry which can connect all buildings with walkable paths on rough terrain, without impacting the terrain at all.

2: Whenever evaluating, I would always prioritize the actual functionality of the settlement over the visual elements. I saw many entries with large impressive structures, or beautiful prefabricated structures which would be laid out around the terrain, but I would always look for things such as the ability to reach each building, the ease of navigation, if the buildings had interiors etc. For me, this is the most challenging aspect of settlement generation, especially when combined with requiring the settlements to be functional within Minecraft. One challenge in making a Minecraft settlement is there really is not much need for large settlements or houses, as the villagers operate relatively independently, with the only requirements being connected houses and mob security. Further, most players will instead opt to build their own base rather than take residence in a settlement. As such, trying to make a settlement that feels like a settlement, but also makes sense within the world of Minecraft is also challenging. I also think one of the  big challenges is to adapt to the terrain where the settlement is created. Minecraft's own villages often have problems adapting to different terrains, and working within a rocky mountain range or dense jungle with the same algorithm is an interesting challenge.

In terms of long-term challenges, or possibly points I think will be difficult to perform with an algorithm, I do think narrative is in a strange position here. As my primary work is in the field of narrative, I would always keep an eye out for elements of narrative, which went beyond just the style of the buildings, but also involved a sort of history or overarching narrative to the city. I would love to see more generators focus on the historical simulation of a city that guides and changes the building choices over time, incorporating eg. damage from mob raids, or shifting architectural norms over time. The challenge too, is to make this evident rather than just looking like parts of the city are incomplete or misgenerated.

3: Many of the challenges I mentioned in the previous entries are evaluated through the judging criteria, meaning that a wide variety of approaches to generation can achieve good scores in one area, even if they score poorly in others. Personally though, the numerical evaluations don't particularly matter to me, and I usually provide them as an afterthought once I've a rough idea how each map places. The scores also jump quite quickly from a naive user (5) to a team of experts and architects over the period of months (6) which is such a bizarre jump that I never quite know how to handle results which are slightly better than naive but nowhere near expert. I have never seen an entry which can compare to the expert Minecraft builds and I'm not even sure if the goal of this generation should be to match these builds, which are often more aesthetic than functional. Personally, I'd be fine with a general commentary and ranking, using the questions as a criteria, but of course I understand the need for more quantitative measures for the purposes of evaluations and comparisons.

4: Most of what I noted I talk about in separate questions! I note that it is typically obvious when a particular entrant clearly had more resources to work upon an entry, or an academic basis in PCG. These were often the entries which scored the highest, too. I cannot think of a surefire way to equalize the playing field for this type of competition, but I'm always intrigued at how the rankings might change if some of the entrants had more resources to work with.

6: I enjoyed exploring each world and guessing at how different algorithms were combined to produce different results. I think to some degree the most I learnt about through judging are how different elements of settlements are prioritized by different entrants and how they combine different PCG techniques to realize this. City, settlement, etc. generation is a well trodden field and not even necessarily relevant to digital creativity (considering eg. entirely functional city generation). In fact, I often don't consider the "creativity" embedded within the generator as critical, unless the competition were more geared towards aesthetic or artistic settlements rather than practical ones. Not to say that there isn't much creative potential or purpose in creating functional cities or in the current state of GDMC, but rather there isn't much judgement which relates to creative elements of the AI (themes, motifs, novelty, etc.).

7: My impression is always that co-creative vs. AI should be treated as two separate challenges, though both are interesting challenges in the context of Minecraft settlements. My main concern may be that the open format, which currently benefits the settlement generation, may be too broad for a co-creative challenge. Essentially, I view a co-creative challenge as two issues, which involves both the evaluation of the creation process itself, and the final result, both in terms of the human collaborator, and the expert judges, which may in themselves differ. The current set up of the challenge has allowed a great breadth in terms of not just algorithms, but  approaches to the definition of a "settlement", and more particularly a ``Minecraft settlement'', looking at visuals and functionality and other aspects. However, this breadth means there is no base standard of a settlement, which works well in the current competition but means that as more features are required it will become harder and harder to gain meaningful evaluation from the results. Will an easy to use co-creative tool that creates simplistic cities be better than a dense, incomprehensible tool which can create a wide array of cities? This could actually be a positive though, in that I think this would also challenge the judges and organizers to consider what a framework would be for evaluating and weighting all the different perspectives this new competition would allow.

\subsection{Sean Butler, Judge}

1: Some surprising or outstanding features appeared to be effects which occurred through the possibly conflicting interactions between the generative algorithm and the existing continuous naturalistic environment. In some cases this conflict resulted in surprising outcomes which through their departure from the normal inspired alternative thinking in the user. In other words the `almost' failure or edge case of a couple of algorithms resulted in creative rationalisation on the part of the player. Where generators strove to control the input from the environment their overall outcomes were less impressive.

2: Of the generators assessed most or all demonstrated some limited `artistic' architectural expression. More `functional' architectural expression was observed in some of the larger buildings clearly rooms provided movement flows into other rooms which had different uses. It was unclear how much of this was learned by the AI or coded in. In some cases some of the smaller buildings had positive aesthetic qualities but submissions didn't manage to combine these with larger or more complex structures.

This could in part be due to the block nature of MC, but buildings on MC (e.g. hypixel, hermitcraft) servers show considerably more decoration, integration with the environment and functional features than those generated during the competition. Overall they felt like they had maybe been produced by children or new players. This in itself is an impressive feat, but there is still a long way to go before they approximate experienced human standard.

3: Simple evaluation functions currently used in CI research are insufficient to evaluate complex aesthetic artifacts like settlements and are likely to remain so for some time. Even those within a game like MC. Human evaluation can be a more refined assessment because access to the wider culture to cultural artifacts, settlements, architecture from outside the domain.

Further refinement of the questions might involve an emphasis of the systematic and historical perspectives of the settlements. While these were highlighted in the accompanying text there wasn't much expression in them from the generators. Of course the downside would be that any emphasis in the questions could have a leading effect on the entries.

4: Particularly in older countries its possible to see the evolution of settlements over time in the accumulated buildings of a settlement. As procedural mechanisms go a historical perspective would leave a traces of their progress as the settlement grows.

6: Efforts to move evaluation of CC/CI systems beyond surrogate models and mathematical approaches into the human judgment realm should be pursued.

In the artifacts judged this year the most interesting outcomes came from interactions between the existing environment and the generated content.    

7: A co-creative approach could be on arrow in a quiver of approaches available to creators and prosumers. Naturally some will want a hands off approach and others will struggle to explain their design goals. We should make tools which can support creativity at all levels.

\subsection{Jean-Baptiste Hervé, GDMC organizer}

The GDMC is a great framework for my research in PCG. First of all, because it has a very unique setting: we have quite a variety of generators that are submitted every year (11 in 2020), and then evaluated by humans. This diversity in the submissions alone could be used in researching different generation techniques, and how they achieve different goals. The human evaluation is also interesting in the sense that it tries to capture various aspect of each settlement generator, which are currently complicated to automatically evaluate.

I would also like to point out that while I have been focusing on the settlements themselves, the GDMC also comes with a bonus challenge, which is the Chronicle Challenge. Each entry can also generate an in-game book that tells the story of the settlement since its creation. For the reasons listed previously (various entries and human rating), it could be a good study case for people interested in computational storytelling.

One aspect of the competition that I personally found interesting, and which opens a large field of research, is that a good settlement can have many forms. In terms of narrative, it can be a stronghold, a modern city, or a village, and all those different entries could end up with the same score. This is even more true due to the sandbox nature of Minecraft, where players can prefer a practical or a coherent settlement. One of my goals in my current research is to think in a very generic way, as those settlements can have a huge range of shapes and themes, while being equivalent in human perception.

However, I do have some concerns. After studying the results of the previous years, I can find a relation between the Aesthetic of a generator and its overall grading. Maybe while rating the Functionality of a settlement, some judges are biased by the look of it. Also, how the in game mechanics are actually taken in account in the judging process? For instance, one of the fastest ways of transportation in Minecraft is to use a boat on ice blocks. If a settlement provided roads paved with ice, should it be considered as a perk, or as an unbelievable feature?

Another aspect of the GDMC that makes it a great case study is Minecraft itself, for several reasons. First of all, research in PCG tends to focus on a bunch of very specific genres. While this is an obvious issue in terms of research scope, I consider that Minecraft has several perks that makes it a great candidate for PCG research, in addition to not being a platformer. First of all, Minecraft has no clear genre, and there is no correct way to play it. Some player like it for its challenges, others for its creativity, or even just as an adventure. Through the years, the Minecraft community has come up with hundreds of ideas to change the game: mods, maps, design, and so on. This massive community could be used for research that would involve human testing. Map creators could be involved in the creation of a GDMC entry, testing/rating of submissions, or even in the conception of mixed-initiative creation tools. Minecraft is also a game that relies heavily on PCG, and generating a settlement on top of a generated map might be an interesting case study.

I would also like to make a couple of suggestions to take the competition further. Maybe the scope of the competition could be extended easily, by offering new possibilities to submission. The current entries are mostly generating settlements by adding and removing blocks. Maybe it would be interesting to provide a framework that lets generators place NPCs, animals, perhaps non-block items, or has the ability to write signs. Still within the framework, maybe a judge could be provided some assisting tools. Examples would include a UI that lets them know if a mob can spawn in the settlement, or a list of placed blocks that cannot be built with the local resources. Finally, while the jury's scoring is interesting data, maybe it would be interesting to have a `public' grading, too. The large player base of Minecraft could be put to use here.

\subsection{Adrian Brightmoore, Advisory Board Member and Competitor}

1: I’ve found the generated settlements to be of good quality when assessed for coherence. There are two approaches taken in the submissions that I have noted: one where the landscape is flattened and homogenised into a platform for the purpose of placing buildings and features, and the other where the landscape is used in-situ without significant modifications. The former tends to evoke a ‘city’ feeling, and the latter feels more like a village.

The village approach interests me more than the city approach since it is clearer that the code solution has been sympathetic to the landscape offered by the challenge. The dimension of the challenge maps also lends itself to a village or town scale and not a city, since the scope and scale of a roughly 300 square meter area can be rapidly exhausted by sky-scrapers as was shown by submissions for 2020.

I have the impression that there are now well solved problems in the challenge and that it would be appropriate to offer a framework in the future that includes pre-built generators for these problems. They include:
\begin{enumerate}
    \item Building site profile and selection (is this flat area big enough? Is it land and not lava?)
    \item Roadways, bridges, and pathing between sites of interest
    \item Placing pre-built structures
    \item Writing entries to a book
\end{enumerate}

2: The approach to settlement generation has often been characterised by the placement of pre-generated template assets. This leaves room for future efforts in the area of interior design and layout. My preference would be to see more procedural algorithmic efforts at all scales of the environment from town layout at the top-most level, all the way down to the contents of chests and furnaces within houses.

3: I think the assessments so far have been based on visual impression and exploration. This is ok because it meets the stated objectives of the challenge: ``The goal is to basically produce an algorithm that can rival the state of the art of what humans can produce'' (GDMC website Jan`21)

\begin{figure}[ht]
  \centering
  \includegraphics[width=\linewidth]{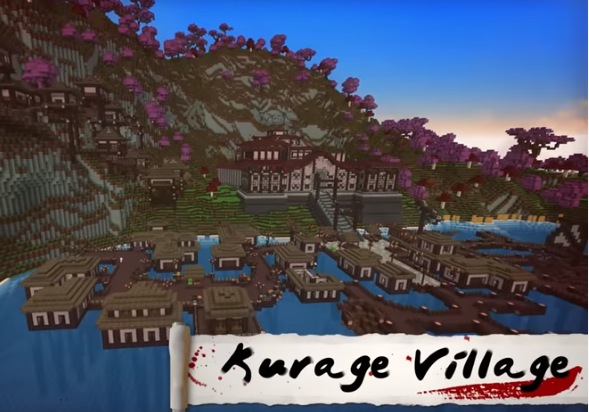}
  \caption{The Land of Akane, by FyreUK. One of the examples presented for the GDMC challenge at   \url{http://gendesignmc.engineering.nyu.edu/rules}}
  \label{fig:fyre}
\end{figure}

Many of the examples given for the challenge are massive themed cities, suggesting that this is the bar to be met by the contestants. Were that the case I think it would be a boring challenge to participate in. I am pleased that the contestants have so far sought to go off in other directions. To expand on this thought, while people use Minecraft as a canvas for their art, the player experience of Minecraft settlements is very very different to the stereotypical homogeneous style of ``Westeros'' castle, similar to the example in Fig.~\ref{fig:fyre}, that is often held up as the peak of build skills.

\begin{figure}[ht]
  \centering
  \includegraphics[width=\linewidth]{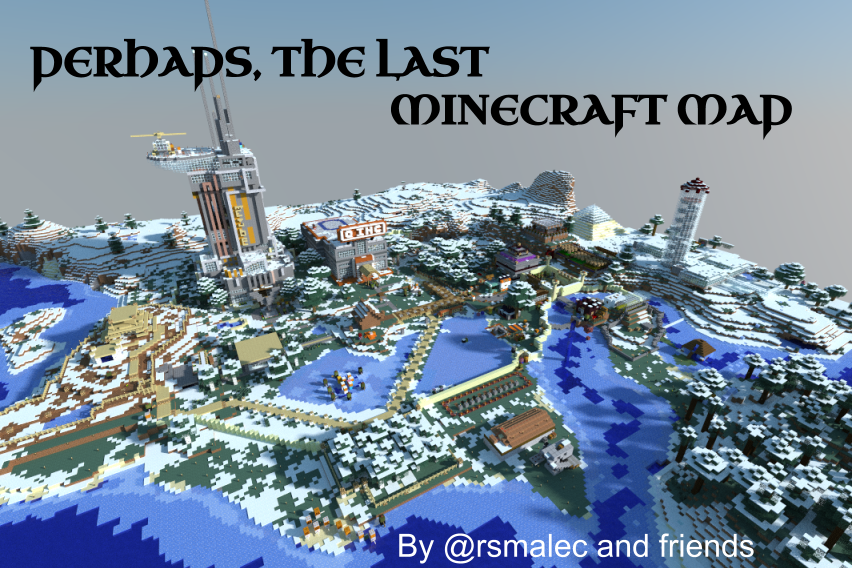}
  \caption{Typical community settlement from the multiplayer server project by Ron Smalec (@rsmalec twitter), characterised by divergent build styles, game-functional elements like mob farms, and disjoint construction agendas that are tied to each individual’s gameplay experience.}
  \label{fig:smalec}
\end{figure}

At the other end of the scale, if we were truly attempting to replicate a human-built settlement from an existing Minecraft server, we’d have a regular grid of building plots with various random styles on it. It wouldn’t be very pleasant to look at, but that’s the reality of how multiplayer settlements evolve (Fig.~\ref{fig:smalec}). 

I also feel there is room in the challenge to mine and assess the generative procedures, instead of just their output, to determine whether the approaches taken by contestants offer interesting insights into the problems and how they may be more fully solved in the future. I find research in this area fascinating and I have found it has been only lightly explored in the challenge assessments so far.


4: There is a tendency to bring more curved and complex shapes into the settlements which is to be commended. 
The use of lighting and the attempts to evaluate submissions in a range of lighting conditions was highly interesting.


5: My 2018 submission attempted to address the challenge of profiling the landscape for suitable building locations. The 2019 submission looked at the procedural generation of buildings and their contents. The 2020 submission was targeted at agent-based generation with variation in materials and pattern. The 2020 submission leveraged the agents to inform and realise the chronicle challenge.

The timeframe is always a challenge because the process of building a generator from scratch erodes the quality of the submission for me. I find that I reach a point where I know the solution is partial and I’m happy with the framework but lack the interest or time to populate it with sufficient assets to build out a truly compelling result.

I think there would be a higher volume of participation if the contestants were offered more options to assemble existing generators as modules that would then allow them to focus instead on their area of interest. This might help accelerate the competition evolution even faster as teams can reduce the scope of their effort to those tasks that truly interest them.

6: GDMC has resulted in a few treaties on the creative process. Not everyone has shared their process. Setting the challenge up as a competition there may be a sense of reluctance on the part of participants to share openly and transparently each year prior to submission. Going forward it may be appropriate to forego the scoring in the interest of increasing the community involvement and participation.

I don’t think that the process of creating is well exposed by the competition as the competition framework does not directly incentivise participants to share insights into their process. This would be addressed going forward by adding a requirement to do so on submission.

My creative process has been documented each year here for 2018\footnote{\url{http://www.brightmoore.net/what-s-happening-now/gdmc2018competition-aparticipantsperspective}} (Fig.~\ref{fig:Brightmoore-2018})
\footnote{\url{https://twitter.com/TheWorldFoundry/status/1013020797152124929?s=20}}
, 2019\footnote{\url{http://www.brightmoore.net/what-s-happening-now/gdmc2019-proceduralsettlementgenerationinminecraftround2}}
\footnote{\url{https://twitter.com/TheWorldFoundry/status/1084804958564757506?s=20}}
and 2020\footnote{\url{https://twitter.com/abrightmoore/status/1275878848974843905?s=20}}. 

My approach to the competition is through hierarchical methods. The settlement is decomposed into building/farm/feature assets, which are decomposed into rooms/doors/windows/contents (for buildings) and other feature-specific generators. At each level of hierarchy the relation between the components is assessed and in some cases connected (with paths, for instance, as in 2018 and 2019). 2020 was somewhat different in that the building strategy was highly random. In this way my 2020 submission was more like the first type of submissions that disregard the terrain entirely. I think this is not a highly creative path to take but I enjoyed the results and the path to get there. Therefore it’s difficult for me to answer this question one way or the other because the computational creativity in 2020 for me was about compensating for randomness; settling on a strategy that looks good and works well without having an overarching strategy.

7: This approach might be more suited to Project Malmo or Computercraft platforms as this is territory they are already well positioned in as far as automaton frameworks are concerned.

For what it’s worth, I think you’re anticipating an ingame automation and I have little interest in that for technical reasons. The scope and scale of what can be achieved by editing the world is far greater than that which can be achieved when constraining the designer to work within an area prescribed by the currently loaded-chunks in game.

I would welcome the GDMC competition expanding the scope to allow more expansive submissions with a scale that allows for the types of cities originally envisioned in the brief back in 2018. Here is a prototype example of what may be possible in time given an expanded scope:

\begin{figure}[ht]
  \centering
  \includegraphics[width=0.49\linewidth]{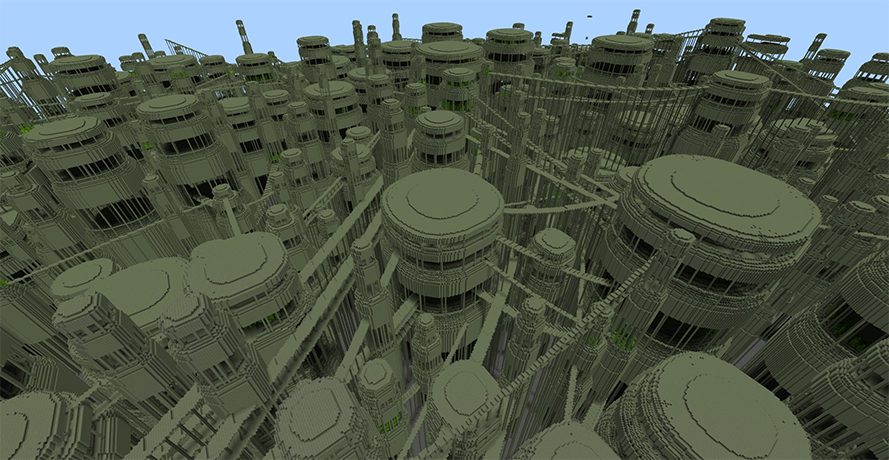}
    \includegraphics[width=0.49\linewidth]{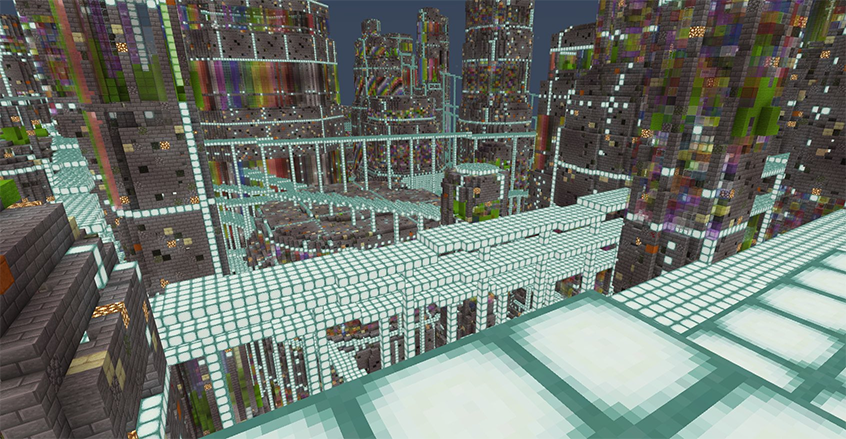}
  \caption{Prototype and realization of large scale generation from 2018. Development notes online.}
  \label{fig:Brightmoore-2018}
\end{figure}

\subsection{David Mason, GDMC Participant}

I feel the main challenge for competitors is to create settlements with a strong aesthetic, and have some variance between buildings in the settlement such that all buildings do not look the same. Contestants must find a happy medium between hard coding many building options and using a complex method of building generation such as a GAN. A significant challenge also appears to be creating detailed interiors past just placing chests and furnaces and such. 

Human judging of the categories does work, however, the familiarity of judge with Minecraft may have an effect on judging as judges very familiar with Minecraft may have a different definition of functionality to someone not familiar with Minecraft as someone not familiar with Minecraft may not be familiar with all the game mechanics that would make a settlement functional. 

When listening to the live impressions with AIandGames it was interesting to see how different his interpretations of the settlements were to what I expected them to be.

There were many challenges creating my generator, the most time consuming of which was ensuring the correct rotations of blocks such as stairs and trapdoors. Additionally, a problem I never solved was placing tile entities properly. For example. placing a chest or a bed would place invisible blocks that a player could interact with but could not see. 

I used to believe that constraints breed creativity, however, this process showed me that for digital design the opposite may be true. As, the constraints that I faced, I felt, did not inspire me to overcome problems in creative ways.

I feel like typically it is rare to generate centerpieces through algorithms, usually they are more hand made. In this sense, this competition is taking interesting, useful steps in digital creativity. The more accessible tools for generative design becomes, the more creativity will be seen, hopefully human-ai co-creativity will help make generative design more accessible. 

The way a human-ai co-creativity tool would work in my mind would be similar to cycleGAN in that you can change the style of something already there. In Minecraft that might look like being able to convert a stone cube into a house. I have no idea if this is achievable or how it would be done. But, I do believe it would be helpful. Anything to relieve the tedious tasks or remove steps between actions in the design process would be helpful e.g. figuring where to place buildings in the settlements.

\subsection{Claus Aranha, Advisor to multiple GDMC teams}


I have supervised students who participated in the 2019 and 2020 editions of the GDMC. As a teacher, I feel that GDMC is a fantastic challenge for intermediate programming students, who have the basics of programming down, and now need a larger scale challenge requiring the design of larger scale programs and actual application of the algorithms they learned in the classroom. I'll never forget the face of one particular student, after I suggested to her how she could use a minimum spanning tree algorithm to decide where to lay down paths in her settlement: "So that's what the graph theory classes are good for!".

The results of all participants are extremely interesting. The buildings are very detailed, and most submissions seem to be very good at putting large amount of buildings at appropriate places in the Minecraft landscape.

Maybe too good, though?

Although the different participants in the recent competitions show a wide variety of approaches and building styles, they seem to have a common point in the tendency of constructing a large, sprawling, centralized build. Something built by a single author in the game's creative mode, where there are no restrictions in the number of blocks used or in the movements abilities of the player. In other words, the current "meta" of GDMC submissions seems to be geared towards visually appealing large scale settlements.

A different direction, that I feel would be an interesting challenge, is to create smaller, cozier places, that look more like the product of regular play sessions of Minecraft: buildings that are limited in scope by the resources available in the map, and the game's restrictions on movement. One of many possible approaches to achieve this idea is to indirectly generate the settlement by simulating the action of building agents. These agents would be embodied in the Minecraft world, meaning that they would have abstracted movement limited by some of the rules of the game (jumping height, etc), and block breaking/placing actions that are limited by the agent's reach and a virtual internal inventory. Using these actions, the agents would aim to achieve objectives such as building shelter, farming resources, and exploring. The simulation of multiple agents could add an aspect of cooperation or competition in the resulting settlement. The log of actions and tasks from the agents could even be woven into the "Chronicle" challenge of the competition.

The effort necessary to build algorithms to guide such agents would be of a level of complexity much above the current GDMC submissions, but I think this is an effort well worth undertaking. Not only it would result in settlements more similar to those that emerge in Minecraft play sessions, it would also help guide the development of agents that could play along with human players. Also, modification to agent rules could generate a wide variety of settlements, with animal, organic, or alien properties (nether fungal growth settlement anyone?). Not to mention that the log of agent actions could be used to generate a time-lapse of a settlement's construction. I'd be very curious to see that!

\subsection{Ruck Thawonmas, Advisory Member of Team ICE{\_}JIT}

Generator submitted in 2020 competition: ICE-JIT\footnote{\url{https://github.com/ice-gdmc2020/ICE_JIT}}

1 : Two points that I found good or was surprised are subterranean utilization and building interiors. They are described below in the following.

\subsubsection*{Subterranean Utilization}

Shown below is how subterranean was utilized by one of the 2020 generators. This kind of idea did not come to my mind when brainstorming ideas for our generator ICE{\_}JIT. It seems that a kind of dungeon generation technique was used. 

\begin{figure}[ht]
  \centering
  \includegraphics[width=0.49\linewidth]{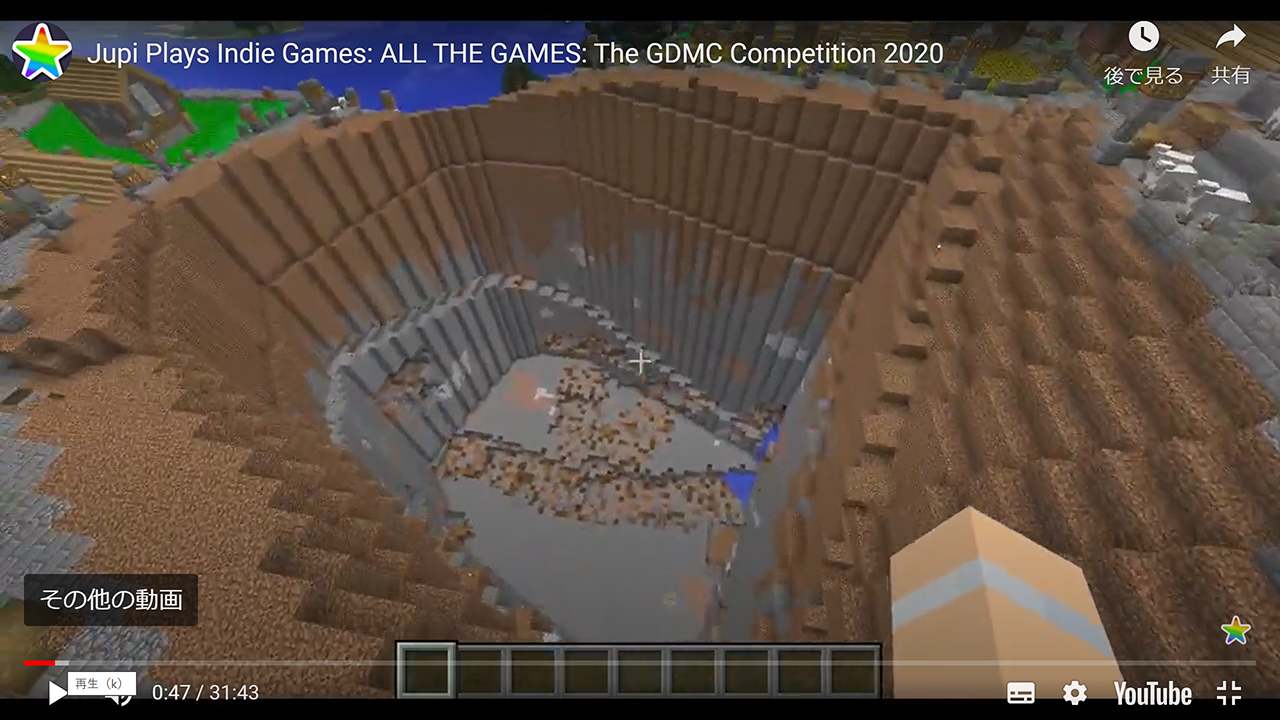}
  \includegraphics[width=0.49\linewidth]{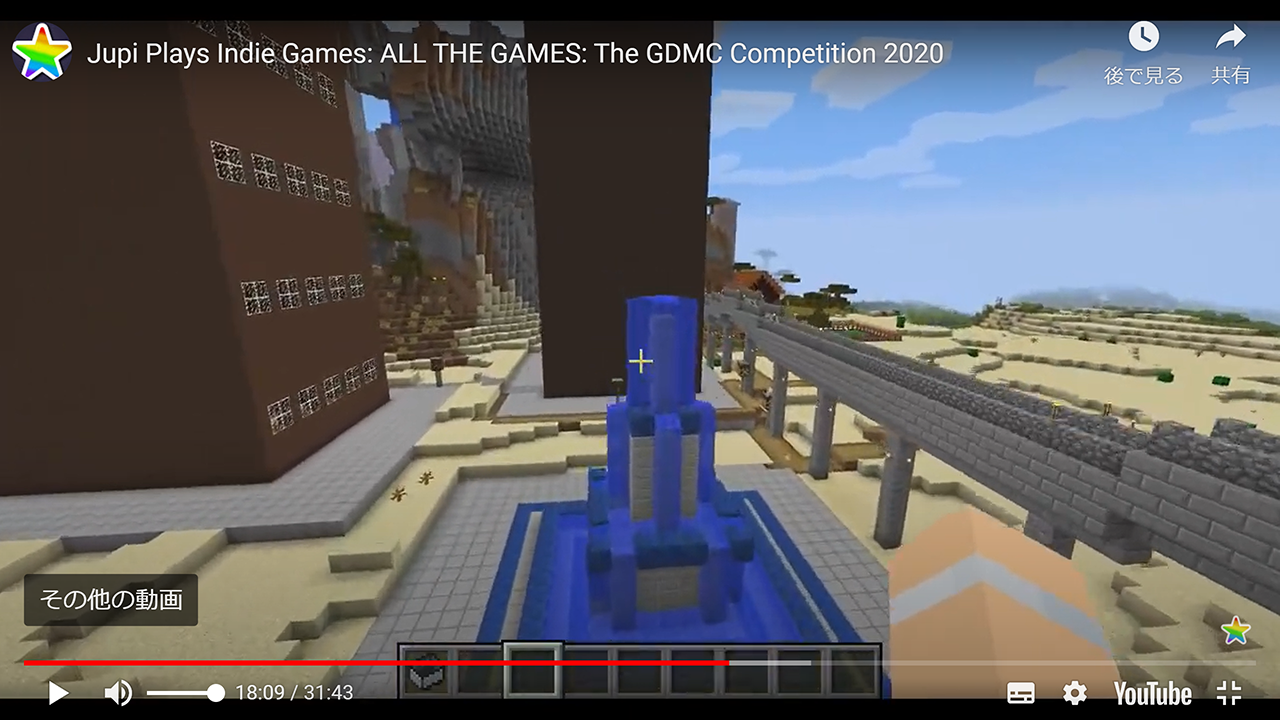}
  \includegraphics[width=0.49\linewidth]{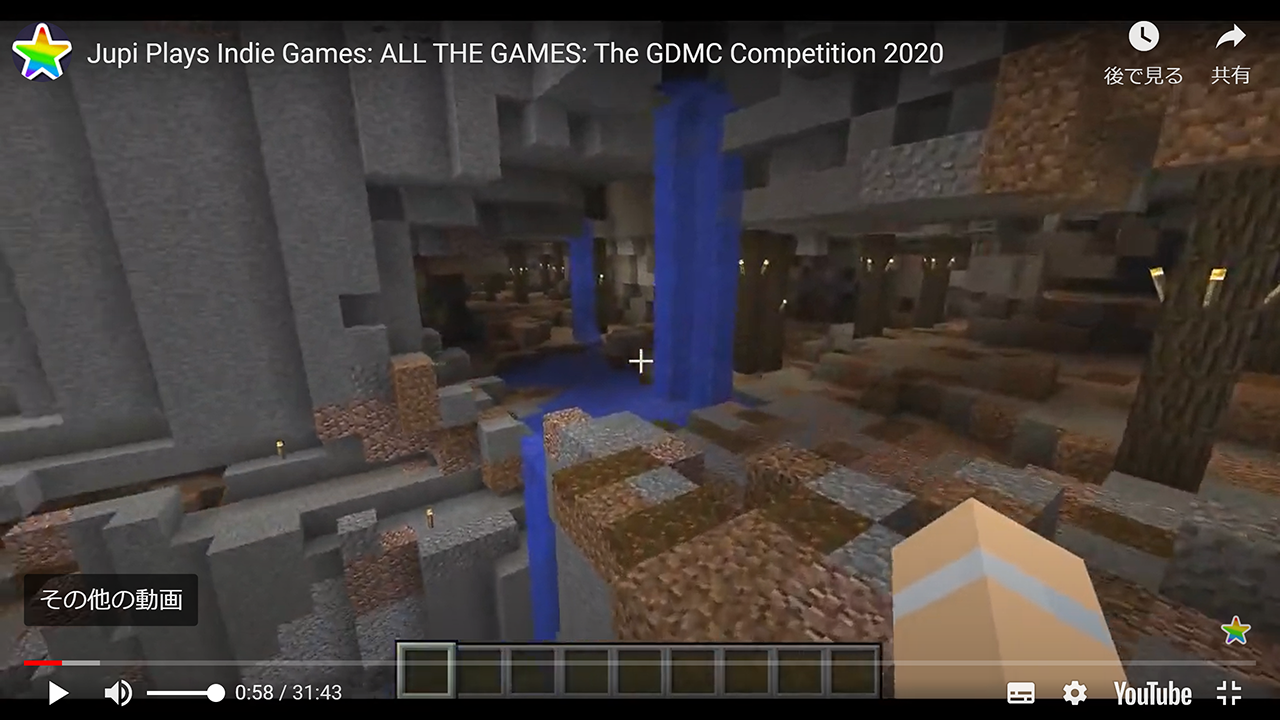}
  \includegraphics[width=0.49\linewidth]{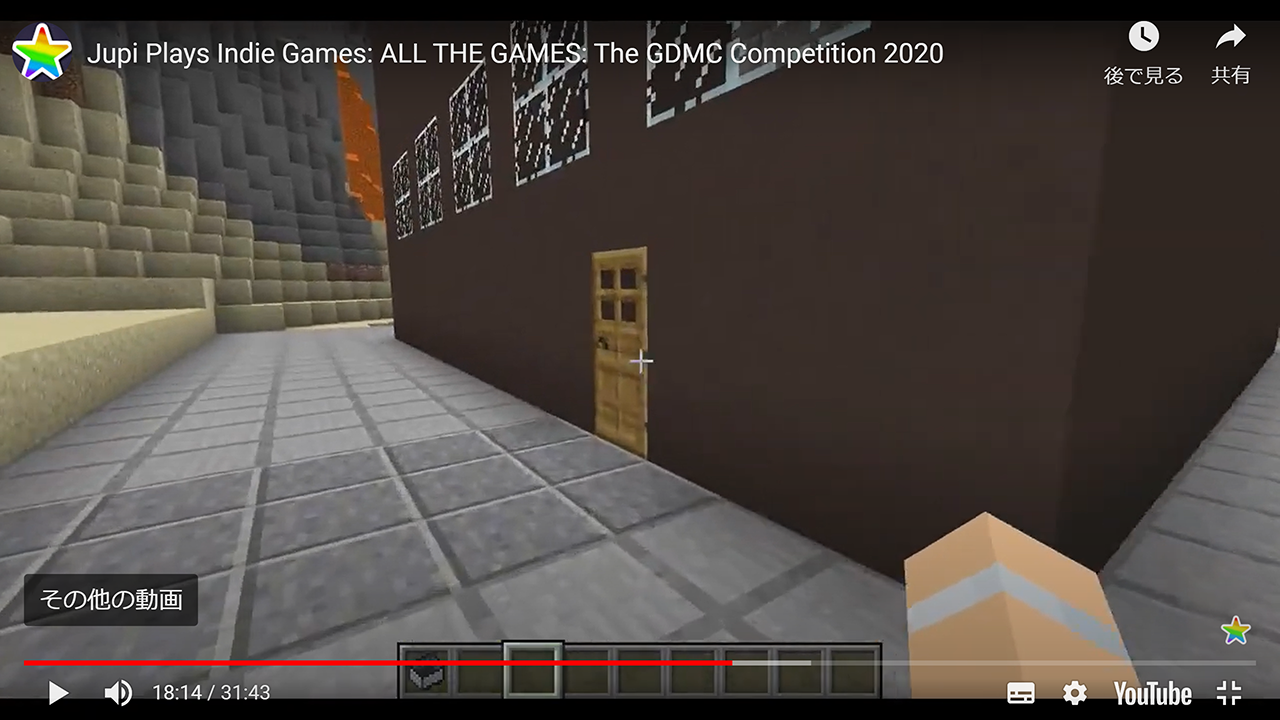}
  \includegraphics[width=0.49\linewidth]{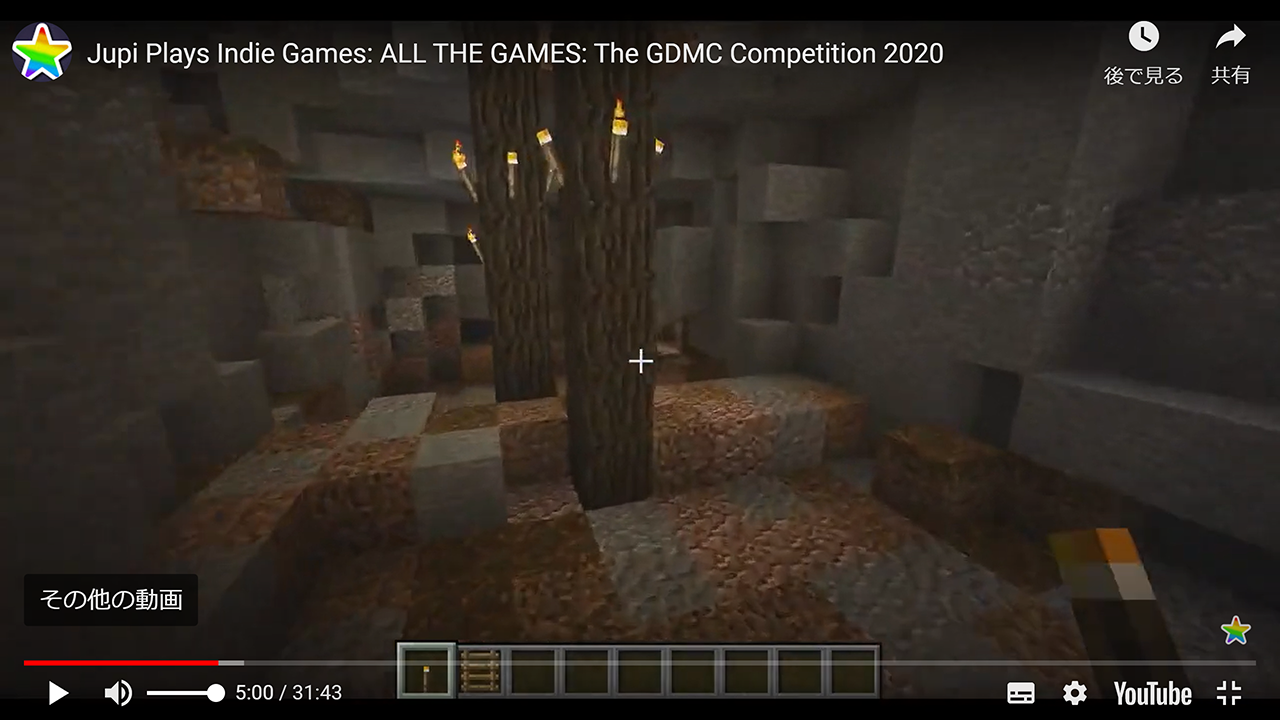}
  \includegraphics[width=0.49\linewidth]{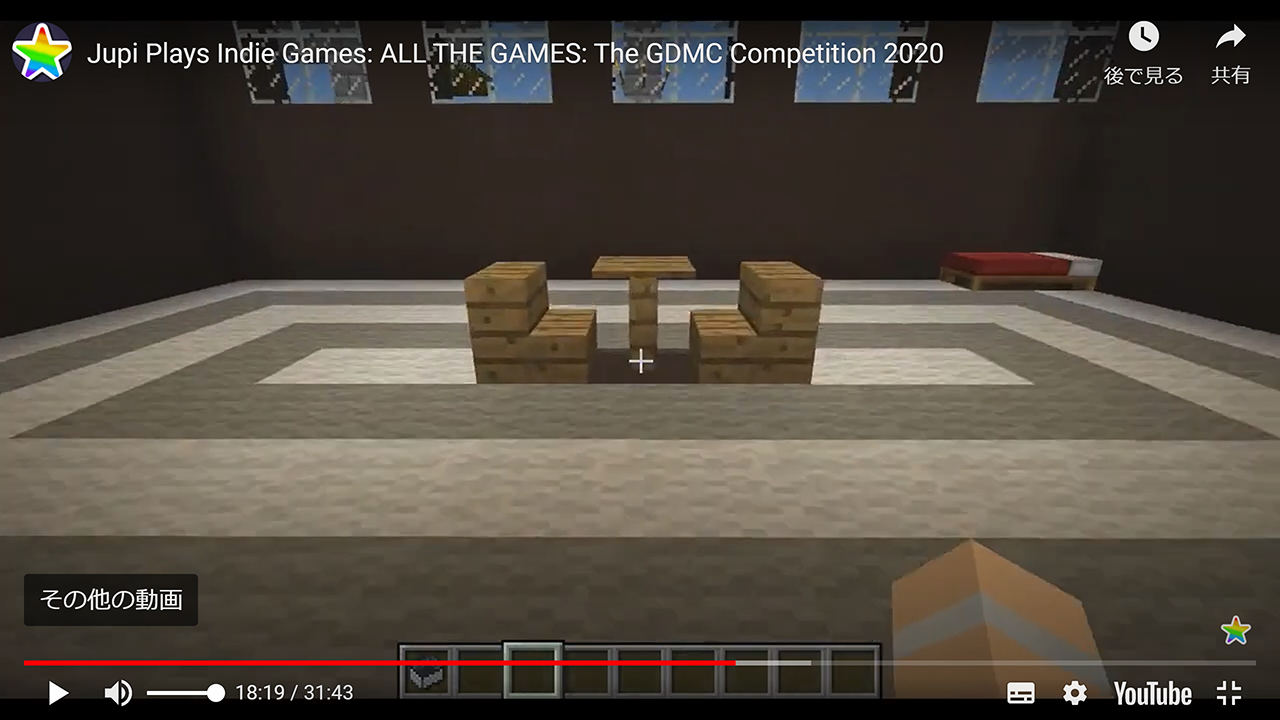}
  \caption{Subterranean Utilization examples (3 images on the left), and Building Interiors examples (3 images on the right). Pictures from (\url{https://www.youtube.com/watch?v=e6e3C8hDNFg})}
\end{figure}

\subsubsection*{Building Interiors} 

Buildings themselves are not that hard to procedurally generate. However, properly placing and lay-outing their interiors (stairs, rooms, furniture, etc.) require some thoughts for implementation. Examples by one of the 2020 generators are shown below. 

%

2: I am not sure how much we can play with physics in Minecraft. However, in case that findings from this competition are to be applied in real world settings, can constraints such as the ability to tolerate (no-collapse under) strong winds/earthquakes be introduced? Introduction of these kinds of constraints would make the competition very challenging. 

3: To me, evaluation by humans is almost OK. However, from YouTube videos uploaded by some judges for the 2020 competition, it seems that the amount of time they spent judging is different. It is not clear either if the judges could explore all the salient features of settlements generated by each generator. A possible solution for this is to have each team additionally submit a video summarizing their settlement on each map. Such videos can be used by judges as their reference on where/what to explore. In addition, to increase the quality of evaluation, auxiliary judges can be introduced by, for example, asking FDG participants, who do not have conflicts of interest with any teams, to vote based on those videos during the conference period.

4: Some settlements from different 2020 generators look pretty much similar. It might be due to that generator developers had no room to focus more on the design part and thus must devote their time in implementing many modules/tools for generation of their settlements. As a result, it might be a good idea if useful modules/tools are provided by the organizing team.

\begin{figure}[ht]
  \centering
  \includegraphics[width=\linewidth]{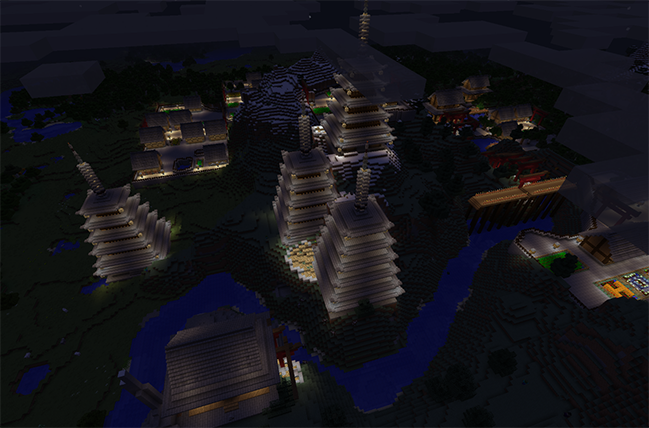}
  \caption{ICE{\_}JIT settlement in Light Up.}
  \label{ref:light}
\end{figure}

5: The main concept behind our generator ICE{\_}JIT is to build unique settlements not seen in previous competitions. Our challenge was in finalizing the idea among many candidates. Eventually we decided to go with a Japanese style. After the concept was fixed, implementation by our student members went smoothly. As an advisory member, When I first saw one of our settlements in Light Up, shown in Fig.~\ref{ref:light}, I was a bit surprised in its beauty, which was beyond expectations.

Our humble advice to others is “Come up with unique ideas that you yourself are thrilled to see – don’t care much about how judges will evaluate them!”

6: Yes, the GDMC competition is doing well toward the aforementioned aim. Since evaluation plays an important role in digital creativity, more ideas about evaluation are elaborated here. Namely, it might be good if we can come up with a metric to evaluate reliable judges or to develop an AI (and reliable) judge. For this line of research, the human judges’ log action/movement data (in the format similar to that of minerl.io) and their evaluation should be made publicly available. A sister (sibling?) competition can also be organized where participants submit their judge AI that should output a ranking similar to the ranking by human (and reliable) judges.  

7: Yes, it’s a good goal. A challenge might lie in the types of users who will be interacting with those AIs. If non experts, in terms of making Minecraft (or real word) objects/architectures, do that, training should be developed and provided to them in advance. In other words, as far as the fairness of the competition is concerned, it is important to ensure that each generator is assigned a set of equal-skill human partners for the targeted co-creativity task. 

\subsection{Mike Preuss and Hagen Fischer, Judges}

On the first glance, the settlements produced during the GDMC
competition 2020 already look quite amazing.
House styles that adapt to different biomes or even houses built in a
believable way on water as in real-world Venice are amazing and invite
to have a closer look at what automated Minecraft building can do
already. Generally, it appears as there is progress, the resulting
villages often look more refined, too unrealistic construction attempts
are more rare.
With respect to the main aim of the competition, to improve algorithmic
design capabilities for complex multi-component designs as settlements,
three single aspects may be of specific interest:

\begin{enumerate}
\item houses or places and all the single constituents of villages have to
be constructed themselves in a meaningful and appealing way,
\item the overall organization of the villages (what types of houses, how
are they layed out) has to look either constructed or sort of randomly
composed, and ideally both to a certain extent,
\item consistency breaches have to be avoided: small mistakes that destroy
the immersion can ruin the whole impression and get otherwise neatly
designed villages rated down.
\end{enumerate}

On the first aspect, the participants did a good job, there is a lot of
detail, and a lot of variance. Sometimes, single parts as e.g. too big
foundations with respect to house sizes seem to be out of balance.
The composition of villages is largely good, and looks improved in
comparison to last years competition entries, but there are some
examples where the whole composition is devalued by an added feature
that is intended to improve the overall impression as e.g. a town wall
that is placed too far apart from the inner area of the village.

Implementing diversity (or rather a mixture of structure and randomness
that may be regarded as organic growth of a settlement) on an existing
terrain is probably the most difficult aspect of the challenge, and the
results we have seen here are of mixed nature.
In some cases, it works great, especially on flat terrain. Roads and
paths in such an environment often look realistic, the fading out of
some paths appears natural. Nevertheless, in some cases, the generators
have their problems to find a good solution. This often occurs  for lush
locations, open caves, high mountains, ravines, or more generally, in
ruff terrain. If a house is buried by sand, this may first promise to be
a nice story but if this house even got an intact field and flowers
which alone can hold tons of sand, something is off. If a big plaza made
out of cobblestone can balance on a 50 meter tall stone pillar or if a
step is 10 meters high and impassable due to the incline of a mountain,
the composition looks so unrealistic that the immersion that is
otherwise created by the village composition breaks.

The effect that the procedurally generated design on the one hand needs
to be as good as possible in terms of performance (here, visual
appearance), but on the other hand needs to be as believable as possible
in order to avoid removing
the suspension of disbelief is not a new insight, however. The BotPrize
\cite{hingston2009turing} was a behavior based Turing test for Unreal
(FPS) bots around a decade earlier than the GDMC. Its best entries after
5 years of competition
\cite{karpov2013believable} had put a lot of effort into imitating
humans and their behavioral patterns, so that human judges were simply
not able any more to tell the difference between a human player and a
bot. For creating a more realistic impression of the GDMC generated
villages from the viewpoint of a human observer, especially if the human
is not a PCG expert, it will be very important in future not only to add
more features, but also to focus on removing the many small "mistakes"
that disturb the overall impression.

\end{document}